\documentclass[10pt,twocolumn,amsmath,amssymb,aps,prb]{revtex4-1}

\usepackage{graphicx}
\usepackage{dcolumn}
\usepackage{bm}
\usepackage{hyperref}
\usepackage[cp1250]{inputenc}
\usepackage[usenames,dvipsnames]{color}
\usepackage{tabularx}
\usepackage{cancel}
\usepackage{color}

\newcommand{\Si}{\mathop{\rm Si}}

\newcommand{\rmi}{{\rm i}}
\newcommand{\rmd}{{\rm d}}

\def\XXint#1#2#3{{\setbox0=\hbox{$#1{#2#3}{\int}$}
     \vcenter{\hbox{$#2#3$}}\kern-.5\wd0}}

\renewcommand{\today}{submitted version as of September 3, 2021}

\begin{document}

\title{Tracing the Mott-Hubbard transition \\
  in one-dimensional Hubbard models without Umklapp scattering}


\author{Florian Gebhard$^1$}
\email{florian.gebhard@physik.uni-marburg.de}
\affiliation{$^1$Fachbereich Physik, Philipps-Universit\"at Marburg,
  35032 Marburg, Germany}
\author{\"Ors Legeza$^{1,2,3}$}
\email{legeza.ors@wigner.hu}
\affiliation{$^2$Strongly Correlated Systems Lend\"ulet Research Group, 
Institute for Solid State Physics and Optics, MTA Wigner Research Centre for
Physics, P.O.\ Box 49, 1525 Budapest, Hungary}
\affiliation{$^3$Institute for Advanced Study,
  Technical University of Munich, Lichtenbergstrasse 2a, 85748 Garching, Germany}

\date{\today}

\begin{abstract}%
We apply the density-matrix renormalization group (DMRG)
method to a one-dimensional Hubbard model
that lacks Umklapp scattering and thus provides an ideal case to study the
Mott-Hubbard transition analytically and numerically.
The model has a linear dispersion and displays a metal-to-insulator transition 
  when the Hubbard interaction~$U$ equals the band width, $U_{\rm c}=W$,
  where the single-particle gap opens linearly, $\Delta(U\geq W)=U-W$.
  The simple nature of the elementary excitations permits to determine numerically
  with high accuracy
  the critical interaction strength and the gap function in the thermodynamic limit.
  The jump discontinuity of the momentum distribution $n_k$ at the Fermi wave number
  $k_{\rm F}=0$  cannot be used to locate accurately $U_{\rm c}$
  from finite-size systems.
  However, the slope of $n_k$ at the band edges, $k_{\rm B}=\pm \pi$,
  reveals the formation of a single-particle bound state which can be used to
  determine $U_{\rm c}$ reliably from $n_k$ using accurate finite-size data.
  \end{abstract}



\maketitle

\section{Introduction}
\label{sec:Intro}

The Mott-Hubbard metal-to-insulator transition poses one of the
fundamental and
most intriguing problems
in condensed-matter many-particle physics.\cite{Mottbuch,Gebhardbook}
When there is on average one electron per lattice site in a single $s$-band,
and the electrons are supposed
to interact only locally with the Hubbard interaction of strength~$U$,
there must be a transition from a metallic state at $U=0$ to an insulating state
at $U>U_{\rm c}$. In generic situations,
the critical interaction strength $U_{\rm c}$ should be of the order
of the bandwidth~$W$ because the Coulomb interaction competes
with the electrons' kinetic energy.
Apparently, the interaction-induced metal-to-insulator transition cannot be
attacked using perturbation theory and thus poses a true many-body problem
that cannot be solved in general even for simple model Hamiltonians such as
the single-band Hubbard model.

Indeed, exact, analytic solutions are scarce and restricted
to one spatial dimension where the physics often is special.
Indeed, the Bethe Ansatz solution~\cite{LiebWu} shows
that the one-band Hubbard model at half band-filling describes an insulator for all finite
interactions. This is the generic situation for one dimensional models
when the two Fermi points in momentum space are separated
by half a reciprocal lattice vector because Umklapp scattering induces
a relevant perturbation at half band-filling
for all $U>0$.\cite{Giamarchi,Solyom3,Kuramotobuch}

Since the induced gaps for single-particle excitations of the half-filled ground state
are exponentially small for small interactions,
it is exceedingly difficult to locate the transition and to calculate the size
of the gap in numerical simulations that are
necessarily restricted to finite chain lengths.

A way out of this dilemma offer modified one-di\-men\-sion\-al models, e.g., those
with only one Fermi point, where
essentially all electrons move in the same direction. An example for such a model
is the $1/r$-Hubbard model
where the dispersion is linear over the whole first Brillouin
zone.\cite{GebhardRuckenstein,Gebhardbook}
The standard Hubbard model and the $1/r$-Hubbard model are limiting cases
of the $1/\sinh(\kappa r)$-Hubbard model with electron transfer range $1/\kappa$.
This model can be solved exactly with the help of the Asymptotic
Bethe Ansatz.\cite{SutherlandABA,SutherlandinShastry}
With only one Fermi point present at $\kappa=0$,
Umklapp scattering is absent, and
the $1/r$-Hubbard model displays the Mott-Hubbard transition at a finite value,
$U_{\rm c}=W$. The single-particle gap opens linearly above the transition,
$\Delta_1(U\geq W)=U-W$. 

In contrast to generic Bethe-Ansatz solvable models, the spectrum
of the $1/r$-Hubbard model is fairly simple and can be expressed in terms
of an effective Hamiltonian for four hard-core bosons that represent
the four possible sites occupations
(Ashkin-Teller model).\cite{GebhardRuckenstein,Gebhardbook}
Consequently, the ground-state energy is a simple sum of ${\cal O}(L)$ terms
where $L$ is the number of lattice sites.
For this reason, the model also provides a perfect testing ground for the development
and test of numerical many-particle techniques
such as the density-matrix renormalization group (DMRG) method.
However, since the electron transfer amplitudes are long-ranged and complex,
standard DMRG codes that are tailored for short-range electron transfers
and interactions are insufficient.

In this work, we study the Mott-Hubbard transition when it is not driven by Umklapp
scattering processes, and present alternative approaches to locate
quantum phase transitions in many-particle systems when conventional
extrapolations, e.g., for the gap, lead to inconclusive results.
We use DMRG to calculate
the ground-state energy and the single-particle and
two-particle gaps with high accuracy. Moreover, the DMRG permits the calculation of
ground-state properties that cannot be accessed through the spectrum, e.g.,
the momentum distribution $n_k(L;U)$, for finite system sizes 
and interaction strengths. 
We monitor the Mott-Hubbard
transition from the single-particle gap
as a function of system size and interaction strength, and
also track the Mott-Hubbard transition from $n_k$,
in the apparent jump discontinuity $q(L;U)$ at the Fermi wave number
and in the slope $s_{-\pi}(L;U)$ of the momentum distribution at the band edge.

The successful analysis
of the Mott-Hubbard transition in the $1/r$-Hubbard model
paves the way for a DMRG study of the
$1/r$-Hubbard model with nearest-neighbor and long-range interactions
which may change the nature of the Mott-Hubbard transition.
We intend to address this latter issue in a forthcoming publication.

Our present work is organized as follows. In Sect.~\ref{sec:modeldef} we define the
model and discuss the ground-state properties of interest, namely,
the ground-state energy, the single-particle gap, the two-particle gap, and the
momentum distribution. In Sect.~\ref{sec:gsenergyandgaps}
we discuss important aspects of our DMRG algorithm and
analyze the finite-size dependence of the ground-state energy and of the gaps.
In Sect.~\ref{sec:momentumdistribution} we present the 
momentum distribution of the $1/r$-Hubbard model from DMRG calculations
with up to $L=128$ sites for various interaction strengths $0\leq U\leq W$,
and compare it
to perturbative results from weak and strong coupling.
In Sect.~\ref{sec:locateMHtransition} we show that the simple spectral structure
of the $1/r$-Hubbard model permits to locate with high accuracy
the critical interaction and the critical exponent for the single-particle gap.
The apparent jump in the momentum distribution does not provide
a good estimate for the transition. However, the slope of the momentum
distribution at the band edge displays a resonance-shape behavior that
indicates the existence of a single-particle bound state at the band edge
in the thermodynamic limit when $U=U_{\rm c}$.
Short conclusions, Sect.~\ref{sec:conclusions}, close our presentation.
The conventional analysis of the finite-size gap data is deferred to appendix~\ref{app:A}.
In appendix~\ref{app:B} we motivate the observation of a Fano resonance structure
in the slope of the momentum distribution at the band edges
as a function of the interaction strength.

\section{Hubbard model with linear dispersion}
\label{sec:modeldef}

\subsection{Hamiltonian}

We address the $1/r$-Hubbard model~\cite{GebhardRuckenstein,Gebhardbook}
\begin{equation}
  \hat{H}=\hat{T}+U\hat{D}
  \label{eq:fullHubbardmodel}
\end{equation}
on a ring with $L$~sites ($L$: even).


In the $1/r$-Hubbard model,
the operator for the kinetic energy~$\hat{T}$ is given by
\begin{eqnarray}
      \hat{T} &=& \sum_{\substack{l,m=1\\
      l\neq m; \sigma}}^{L}t(l-m)
    \hat{c}_{l,\sigma}^+\hat{c}_{m,\sigma}^{\vphantom{+}} \; ,
      \label{eq:defT} \\
  t(r) &=& (-\rmi t ) \frac{(-1)^r}{d(r)} \; ,\nonumber \\
  d(r)& =& \frac{L}{\pi}\sin\left(\frac{\pi r}{L}\right) \; .
  \label{eq:defTconstituents}
\end{eqnarray}
The creation and annihilation operators $\hat{c}_{l,\sigma}^{+}$,
  $\hat{c}_{l,\sigma}^{\vphantom{+}}$
  for an electron with spin
$\sigma=\uparrow,\downarrow$ on lattice site~$l$ obey the usual
anti-commutation relations for fermions.

In eq.~(\ref{eq:defTconstituents}),
$d(l-m)$ is the cord distance between the sites~$l$ and $m$ on a ring.
In the thermodynamic limit and for $|l-m|\ll L $ fixed,
we have $d(l-m)= (l-m)+ {\cal O}(1/L^2)$, and the electron
transfer amplitude between two sites decays inversely proportional
to their distance (`$1/r$-Hubbard model').

Since $L$ is even,
we have anti-periodic electron transfer amplitudes
because $d(L+ r)= -d(r)$. Therefore, we must choose
anti-periodic boundary conditions
\begin{equation}
\hat{c}_{L+l,\sigma}=-\hat{c}_{l,\sigma}
\end{equation}
for the operators, too.
With these boundary conditions,
the kinetic energy operator is diagonal in Fourier space,
\begin{eqnarray}
  \hat{C}_{k,\sigma}^+
  &= & \frac{1}{\sqrt{L}}\sum_{l= 1}^{L}
  e^{\rmi  kl } \hat{c}_{l,\sigma}^+ \; , \nonumber \\
  \hat{c}_{l,\sigma}^{+}
    &= & \frac{1}{\sqrt{L}}\sum_k
    e^{-\rmi  kl } \hat{C}_{k,\sigma}^+ \; ,    \nonumber \\
    k&=& \frac{(2m+ 1)\pi}{L}\;, \; m= -\frac{L}{2}, \ldots, \frac{L}{2}-1\; ,
    \label{eq:FTofoperators}
\end{eqnarray}
so that
\begin{equation}
  \hat{T}=\sum_{k,\sigma}\epsilon(k) 
  \hat{C}_{k,\sigma}^+\hat{C}_{k,\sigma}^{\vphantom{+}}\; , \quad
\epsilon(k)=t k \; .
\end{equation}
The dispersion relation of the $1/r$-Hubbard model is linear. We set
\begin{equation}
t=\frac{1}{2\pi}
\end{equation}
so that the bandwidth is unity, $W\equiv 1$.


The on-site (Hubbard) interaction~\cite{Hubbard1963,Gutzwiller1963,Kanamori}
acts locally between two electrons with opposite spins,
\begin{equation}
  \hat{D}= \sum_{l=1}^L \hat{n}_{l,\uparrow}\hat{n}_{l,\downarrow}
  \; ,  \quad
\hat{n}_{l,\sigma}=\hat{c}_{l,\sigma}^+\hat{c}_{l,\sigma}^{\vphantom{+}}
  \; ,
\end{equation}
where $\hat{n}_{l,\sigma}$
counts the number of electrons with spin $\sigma$ on site~$l$.


Under the particle-hole transformation
\begin{equation}
  \hat{c}_{l,\sigma}^{\vphantom{+}} \mapsto \hat{c}_{l,\sigma}^+ \quad
  , \quad \hat{n}_{l,\sigma} \mapsto 1-\hat{n}_{l,\sigma} \; ,
\end{equation}
the kinetic energy remains unchanged,
\begin{eqnarray}
\hat{T} &\mapsto& \sum_{\substack{l,m=1\\      l\neq m; \sigma}}^{L}
t(l-m)\hat{c}_{l,\sigma}^{\vphantom{+}}\hat{c}_{m,\sigma}^+
\nonumber \\
&=&
 \sum_{\substack{l,m=1\\      l\neq m; \sigma}}^{L}
  \left[ -t(m-l)\right]
  \hat{c}_{l,\sigma}^+\hat{c}_{m,\sigma}^{\vphantom{+}}
  = \hat{T}
\end{eqnarray}
because $t(-r)=-t(r)$.

Furthermore, the operator for the double occupancy transforms as
\begin{equation}
  \hat{D}\mapsto \sum_{l=1}^L
  (1-\hat{n}_{l,\uparrow})(1-\hat{n}_{l,\downarrow})
  =\hat{D}-\hat{N}+L \; .
\end{equation}
Therefore, $\hat{H}(N_{\uparrow},N_{\downarrow})$ has the same spectrum as
$\hat{H}(L-N_{\uparrow},L-N_{\downarrow})-U(2L-N)+LU$,
where $N=N_{\uparrow}+N_{\downarrow}$.

\subsection{Ground-state properties}
\label{subsec:gsprop}

We are interested in the Mott-Hubbard transition.
The transition can be inferred from the single-particle and two-particle gaps and
from the momentum distribution.

\subsubsection{Ground-state energy and single-particle gap}

We denote the ground-state energy by
\begin{equation}
E_0(N,L;U)= \langle \Psi_0 | \hat{H} |\Psi_0 \rangle
\end{equation}
for given particle number~$N$, system size~$L$, and interaction parameters $U$.
Here, $|\Psi_0\rangle$ is the normalized ground state of the
Hamiltonian~(\ref{eq:fullHubbardmodel}).
We are interested in the thermodynamic limit,
$N, L\to \infty$ with $n=N/L$ fixed.
We denote the ground-state energy
per site and its extrapolated value by
\begin{eqnarray}
  e_0(N,L;U)&=& \frac{1}{L} E_0(N,L;U) \;, \nonumber \\
  e_0(n;U)&= & \lim_{L\to \infty} e_0(N,L;U) \; ,
\end{eqnarray}
respectively.

The single-particle gap is defined by
\begin{equation}
\Delta_1(L;U) = \mu_1^+(L;U)-\mu_1^-(L;U) \; ,
\end{equation}
where
\begin{eqnarray}
  \mu_1^-(L;U)&=& E_0(L,L;U)- E_0(L-1,L;U)
  \nonumber \; , \\
  \mu_1^+(L;U)  &=& E_0(L+ 1,L;U)- E_0(L,L;U)
\end{eqnarray}
are the chemical potentials for adding the last particle to half filling
and the first particle beyond half filling, respectively.
Due to particle-hole symmetry, we have
\begin{equation}
  \mu_1^+(L;U)= U-\mu_1^-(L;U)
\end{equation}
so that
\begin{equation}
  \Delta_1(L;U) = U-2\mu_1^-(L;U)
  \label{eq:defDelta}
\end{equation}
and
\begin{equation}
  \Delta_1(U) = \lim_{L\to\infty} \Delta_1(L;U)
  \label{eq:defDeltaTDlim}
\end{equation}
in the thermodynamic limit.

For finite system sizes, the single-particle gap is always finite,
$\Delta_1(L;U)>0$, due to the discreetness of the kinetic energy spectrum.
When extrapolated to the thermodynamic limit,
the gap $\Delta_1(U)$ vanishes in the metallic phase but remains finite
in the insulating phase. The limiting cases are
\begin{eqnarray}
  \Delta_1(U\ll W)&= & 0 \; , \nonumber \\
  \Delta_1(U\gg W) &= & U-W \; .
\end{eqnarray}
The latter relation can readily be obtained from
strong-coupling perturbation theory.\cite{Gebhardbook}
Thus, the single-particle gap permits to locate the critical
interaction strength for the Mott-Hubbard transition.

\subsubsection{Two-particle gap and effective two-particle repulsion}

Analogously, the two-particle gap is defined by
\begin{equation}
\Delta_2(L;U) = \mu_2^+(L;U)-\mu_2^-(L;U) \; ,
\end{equation}
where
\begin{eqnarray}
  \mu_2^-(L;U)&=& E_0(L,L;U)- E_0(L-2,L;U)
  \nonumber \; , \\
  \mu_2^+(L;U)  &=& E_0(L+ 2,L;U)- E_0(L,L;U)
\end{eqnarray}
are the chemical potentials for adding the last two particles to half filling
and the first two particles beyond half filling, respectively.
We always consider the spin symmetry $S=S^z=0$.
Due to particle-hole symmetry, we have
\begin{equation}
  \mu_2^+(L;U)= 2U-\mu_2^-(L;U)
\end{equation}
so that
\begin{equation}
  \Delta_2(L;U) = 2U-2\mu_2^-(L;U)
  \label{eq:defDelta2}
\end{equation}
and
\begin{equation}
  \Delta_2(U) = \lim_{L\to\infty} \Delta_2(L;U)
  \label{eq:defDeltaTDlim2}
\end{equation}
in the thermodynamic limit.

The two added particles repel each other so that, in the thermodynamic limit,
they are infinitely separated from each other. Therefore, we will have
\begin{equation}
\Delta_2(U)=2 \Delta_1 (U) \; .
\end{equation}
For finite systems, we expect the interaction energy
\begin{equation}
  e_{\rm R}(L;U) = \Delta_2(L;U)-2 \Delta_1 (L;U)={\cal O}(1/L)>0
  \label{eq:defdeltaeR}
\end{equation}
to be positive, of the order $1/L$.
  
\subsubsection{Momentum distribution}

We also study the spin-summed momentum distribution in the ground state
at half band-filling, $N=L$,
\begin{equation}
  n_k(L;U) = \langle \Psi_0 | \hat{n}_{k,\uparrow} + \hat{n}_{k,\downarrow}
  |\Psi_0\rangle 
  \label{eq:defmomentumdistribution}
\end{equation}
with $\hat{n}_{k,\sigma}=\hat{C}_{k,\sigma}^+\hat{C}_{k,\sigma}^{\vphantom{+}}$.
In the metallic phase, the $1/r$-Hubbard model can be classified
as a pure $g_4$-model
within the $g$-ology scheme.\cite{Mahan,Giamarchi,Solyom3,Kuramotobuch}
For this reason, it displays
a jump discontinuity at the Fermi energy $E_{\rm F}= 0$
with wave vector $k_{\rm F}= 0$
(`non-interacting', or `free', Luttinger liquid\cite{KuramotoYokoyama})
in contrast to regular Luttinger liquids that
display algebraic singularities
at $k_{\rm F}$.\cite{MedenSchoenhammer1993,Voit1993,Giamarchi}

In the insulating phase, $n_k(U)$ is a continuous function of $k$
within the first Brillouin zone, $-\pi <k <\pi$.
The limiting cases thus are
\begin{equation}
  n_k(L;U= 0)=  \begin{cases}
    \begin{array}{@{}lll@{}}
    2 & \text{for} & -\pi <k<0\\
    0 &\text{for} & 0<k<\pi
    \end{array}
  \end{cases}
\end{equation}
and $n_k(L;U\to\infty) =  1$.
The jump discontinuity in $n_k(U)= \lim_{L\to\infty} n_k(L;U)$
at the Fermi energy vanishes at the Mott-Hubbard transition.
The discontinuity
may thus be used to located the critical interaction strength.

\section{Ground-state energy and gaps}
\label{sec:gsenergyandgaps}

In this section we compile some analytic results for the
$1/r$-Hubbard model whose spectrum
was conjectured to be identical to that of
an effective Hamiltonian for hard-core
bosons.\cite{GebhardRuckenstein,Gebhardbook}
Therefore, the exact ground-state energy, the
single-particle gap, and the two-particle gap are known for all system sizes~$L$.

These analytic results are accurately reproduced by DMRG
for up to $L=128$ lattice sites. This confirms the validity
of the conjectured effective Hamiltonian.\cite{GebhardRuckenstein,Gebhardbook}
Moreover, it demonstrates the efficiency of the employed DMRG code
for complex-valued, long-range electron transfer amplitudes.

\subsection{DMRG method}

We apply the real-space DMRG
algorithm~\cite{White-1992b,White-1993,Schollwock-2005}
to the Hamiltonian~(\ref{eq:fullHubbardmodel}).
Complex-valued and long-range electron transfer amplitudes
and anti-periodic boundary conditions require
an elaborate DMRG code that was 
originally designed for calculations
in quantum chemistry utilizing various optimization protocols
based on quantum information theory.\cite{Szalay-2015b}

The model has a gapless energy spectrum up to a critical Coulomb
coupling. Therefore, a thorough control of the numerical accuracy
is crucial to obtain accurate values for the gap and for
static single-particle correlation functions.
We make use of the SU(2) spin
symmetry~\cite{McCulloch2007,TothLegeza2008}
and of the dynamic block-state selection
approach (DBSS),\cite{Legeza2003,Legeza2004} where the a-priori value for the
truncation errors was set to $\delta\varepsilon_{\rm Tr}=10^{-6}$ for $L\leq 128$.
The maximal number of selected SU(2) multiplets
according to this accuracy demand turns
out to be around $M_{\rm SU(2)}=4000 \ldots 5000$,
corresponding to about $M_{\rm U(1)}>10000$
DMRG block states when only the total spin in $z$-direction 
is taken into account. We use between seven and eleven DMRG sweeps.

When we compare our DMRG data with the exact results for the ground-state
energies at finite system size~$L\leq 128$ and interaction strength~$U\leq 2W$,
we obtain an absolute error of
$\Delta E_0(N,L;U)=E_0^{\rm DRMG}(N,L;U)-E_0(N,L;U)\lesssim 10^{-4}$
in the energy of the ground state at half band-filling, $N=L$,
and with one or two extra particles
(or holes) in the half-filled ground state, $N=L\pm 1$ and $N=L\pm 2$, respectively.
We used both Davidson and L\'anczos algorithms
as subroutines for the matrix diagonalization. We found the L\'anczos algorithm
to be more stable in all DMRG runs.
As 
tests for the SU(2) and U(1) algorithms we numerically reproduced 
the analytic data for the ground-state energy at half band-filling
with at least six digits accuracy for $U=0.5$, $U=1$, and $U=2$.

We determine the momentum distribution 
from the Fourier transformation of the single-particle density matrix in position space,
\begin{equation}
  n_{k,\sigma}= \frac{1}{L} \sum_{i,j} e^{\rmi k(i-j)} 
  \langle \Psi_0 | \hat{c}_{i,\sigma}^+\hat{c}_{j,\sigma}^{\vphantom{+}}|\Psi_0\rangle \; .
  \end{equation}
The finite-size scaling analysis is carried out for system sizes up to $L=128$
lattice sites. Note that enforced the anti-periodic boundary conditions
lead to a faster convergence of the ground-state expectation values
as a function of inverse system size than in the case of open boundary conditions.
Roughly speaking, the system size must be a factor of two
larger for open boundary conditions than for (anti-)periodic
boundary conditions to obtain the same magnitude for the finite-size corrections.

\subsection{Ground-state energy}

For all system sizes and particle numbers,
the spectrum of the $1/r$-Hubbard model with on-site interactions
and anti-periodic boundary conditions can be obtained from
the hard-core boson Hamiltonian~\cite{GebhardRuckenstein}
\begin{eqnarray}
  {\cal H} &=& \sum_{K}
  h_{K,\uparrow}^s n_{K,\uparrow}^s
  + h_{K,\downarrow}^s n_{K,\downarrow}^s
  +  h_{K}^d n_{K}^d+  h_{K}^e n_{K}^e \nonumber \\
  && 
  + \sum_{K}
  J_K \left[ n_{K-\Delta}^dn_{K}^e-n_{K-\Delta,\uparrow}^sn_{K,\downarrow}^s\right]
\; , \label{eq:two} \\
h_{K,\sigma}& =& \frac{t K}{2} \; , \nonumber \\
h_{K}^e&=&-\frac{t K}{2} \; , \nonumber \\
h_{K}^d&=&U-\frac{t K}{2} \; , \nonumber \\
J_K&=& \frac{t(2K-\Delta) -U +\sqrt{W^2+U^2-2tU(2K-\Delta)}}{2}
  \; .\nonumber 
\end{eqnarray}
In eq.~(\ref{eq:two}) we have
\begin{equation}
  K=\frac{\pi}{L} \left(2 m_K+1\right)  \; , \;
  m_K=-\frac{L}{2}, \ldots, \frac{L}{2}-1 \; ,
  \; \Delta=\frac{2\pi}{L} \; .
\end{equation}
Note that every `site'~$K$ is occupied with either of the four bosons
$\{ \uparrow, \downarrow, e\equiv \circ, d\equiv \uparrow\downarrow\}$.

In the boson language, the ground state is represented by
\begin{equation}
  |\Psi_0\rangle = | \fbox{$\uparrow, \downarrow$},
\fbox{$\uparrow, \downarrow$},
  \ldots \fbox{$\uparrow, \downarrow$}, \circ, \circ, \ldots\circ \rangle
\end{equation}
when $N$ is even.
The first spin is at $K=-\pi+\pi/L$, the last spin is at
\begin{equation}
K_{\rm F}= \frac{\pi}{L} \left(N-\frac{L}{2}-1\right) \; .
\end{equation}
The ground-state energy is thus given by
\begin{equation}
  E_0(N,L;U) = \sum_{K\leq K_{\rm F}} t K
  -\sum_{l=1}^{N/2}  J_{K=-\pi+3\pi/L+2\Delta(l-1)}
  \label{eq:gsenergyfinite}
\end{equation}
where we use that $\sum_K K=0$.
The expression for the ground-state energy per site can be simplified to
\begin{eqnarray}
  e_0 &=&\frac{1}{4} n(n-1) +\frac{U}{4} n\nonumber \\
  && -\frac{1}{2L} \sum_{r=0}^{(N/2)-1}\sqrt{1+U^2-4U(2r+1-L/2)/L}
  \nonumber \\
  \label{eq:gsenergyNeven}
\end{eqnarray}
with $n=N/L$ and $e_0\equiv e_0(N,L;U)$.

In the thermodynamic limit, we find
\begin{eqnarray}
  e_0(n;U) 
    &=& \frac{1}{4}n(n-1)+\frac{U}{4}n \nonumber \\
&&    -\frac{1}{24U}\left[
      (1+U)^3-\left((1+U)^2-4Un\right)^{3/2} \right]\nonumber\\
    \label{eq:ezeroTDL}
\end{eqnarray}
for the ground-state energy per site for $n=N/L\leq 1$,
with corrections of the order $1/L^2$ for $U\neq W= 1$. At the
Mott transition point, $U=W$, the finite-size corrections
are of the order $1/L^{3/2}$.

Table~\ref{tab:one} gives the ground-state energy for various system sizes and
values $U=W/2,W,2W$. The DMRG reproduces the values
with an accuracy of at least six~digits.
On the one hand,
this confirms the validity of the effective hard-core boson model
for system sizes up to $L=128$. On the other hand, it demonstrates the
accuracy and efficiency of the DMRG code for Hubbard models
with complex-valued long-range electron transfers.

\begin{table}[t]
  \begin{center}
  \begin{tabular}[t]{|r|r|r|r|}\hline
    $L$ & $-e_0(0.5)$ & $-e_0(1)$ & $-e_0(2)$  \\
    \hline
    4\vphantom{\Large $A^A$} &  $0.148612632415$
    &    $0.0915063509461$& $0.0472252648292$\\
    6 &  $0.147186159589$
    &    $0.0880376668043$& $0.0443723191770$\\
    8 &  $0.146629075934$
    &    $0.0864886985031$& $0.0432581518589$\\
    16 &  $0.146044361095$
    &     $0.0845160860650$& $0.0420887221900$\\
    32 &  $0.145887166056$
    &     $0.0837683378403$&
          $0.0417743321119$ \\
    64 &  $0.145846869387$
    &     $0.0834913432390$& $0.0416937387731$ \\
    128 & $0.145836722450$
    &     $0.0833902516076$& $0.0416734449005$ \\  
    $\infty$ & $0.145833333333$
    &     $0.0833333333333$& $0.0416666666667$
    \\  \hline
  \end{tabular}
  \end{center}
  \caption{Ground-state energy per site $e_0(L,L;U)$
    of the $1/r$-Hubbard model
    with anti-periodic boundary conditions for various values of $U$
    and $N=L=4,6,8,16,32,64,128$ (half band filling).
    The DMRG reproduces the data with an accuracy of at
    least six digits.\label{tab:one}}
\end{table}

\subsection{Single-particle gap}

For the calculation of $E_0(L-1,L;U)$ we need the ground state
for an odd number of particles, say with $S^z=1/2$.
In the bosonic representation it is given by
\begin{equation}
  |\Psi_0\rangle = | \uparrow, \fbox{$\uparrow, \downarrow$},
  \fbox{$\uparrow, \downarrow$},
  \ldots \fbox{$\uparrow, \downarrow$}, \circ \rangle \; .
\end{equation}
It has the energy
\begin{equation}
E_0(L-1,L;U)= -t K_m -\sum_{l=1}^{L/2-1}J_{K=-\pi+3\pi/L+\Delta+2\Delta (l-1)} \; ,
\end{equation}
where we used that $\sum_K K =0$ and $K_m=\pi-\pi/L$.

This can be simplified to
\begin{eqnarray}
  E_0^-&=&\frac{UL}{4} -\frac{U+1}{2}+\frac{1}{2L} \\
&&  -\frac{1}{2}\sum_{r=0}^{(L/2)-2}  \sqrt{1+U^2-4U(2r+2-L/2)/L}\; , \nonumber 
\end{eqnarray}
where we used the abbreviation $E_0^-\equiv E_0(L-1,L;U)$.
The single-particle gap becomes
\begin{eqnarray}
  \Delta_1(L;U)&=& -1 + \frac{1}{L} \label{eq:DelatfintieOoverrHM}\\
  &&-
  \sum_{r=0}^{(L/2)-2}  \sqrt{1+U^2-4U(2r+2-L/2)/L} \nonumber \\
  && +\sum_{r=0}^{(L/2)-1}  \sqrt{1+U^2-4U(2r+1-L/2)/L} \; .
  \nonumber 
\end{eqnarray}
In the thermodynamic limit, we may use the Euler-MacLaurin sum formula
for the sums in eq.~(\ref{eq:DelatfintieOoverrHM})
to find ($W\equiv 1$ is the bandwidth)
\begin{eqnarray}
  \Delta_1(U)&=&
  \frac{U-W}{2}+\frac{|U-W|}{2} \nonumber \\[3pt]
  &=&\left\{
  \begin{array}{@{}lll@{}}
    0 & \hbox{for} & U\leq U_{\rm c}=W\\[6pt]
    U-W & \hbox{for}& U \geq U_{\rm c}=W
  \end{array}\right.
  \label{eq:gapTDL}
\end{eqnarray}
in the thermodynamic limit. The gap opens linearly at $U_{\rm c}=W$.
The same result can also be obtained from the very definition of $\mu_1^-$.
We use eq.~(\ref{eq:ezeroTDL}) for the ground-state energy for $n=N/L\leq 1$
and find
\begin{equation}
  \mu_1^-=\left.\frac{\partial e_0(n;U)}{\partial n}\right|_{n=1}
  =\frac{W+U}{4}-\frac{|U-W|}{4} 
\end{equation}
which also leads to equation~(\ref{eq:gapTDL}) for the single-particle gap
when we use eq.~(\ref{eq:defDelta}).

\begin{table}[t]
  \begin{center}
  \begin{tabular}[t]{|r|l|l|l|}\hline
    $L$ & $\Delta_1(0.5)$ & $\Delta_1(1)$
    & $\Delta_1(2)$  \\
    \hline
    4\vphantom{\Large $A^A$}  & 0.320867070567
    & 0.567837245196 & 1.39173414113\\
    6 & 0.217167734761
    & 0.435424968083 & 1.26766880285\\
    8 & 0.164130166884
    & 0.362554806107  &1.20326033377\\
    16 &0.082919292022
    & 0.236931208089  &1.10333858404\\
    32 &0.041608809524
    & 0.157825321710   &1.05196761904\\
    64 & 0.020825835837
    &0.106745012744      &1.02602667167\\
    128 & 0.010415720145
    &  0.073053055090   &1.01301894029
    \\
 $\infty$ & 0 & 0 & 1\\
    \hline
  \end{tabular}
  \end{center}
  \caption{Single-particle gap $\Delta_1(U)$
    of the $1/r$-Hubbard model
    with anti-periodic boundary conditions for $U/W=0.5,1,2$
    and system sizes $L=4,6,8,16,32,64,128$
    at half band-filling.
    The DMRG reproduces the data with an accuracy of at least six
    digits.\label{tab:two}}
\end{table}

Table~\ref{tab:two} gives the single-particle gap for various system sizes and
values $U=W/2,W,2W$. The DMRG reproduces the values
with an accuracy of at least six digits.
Again, these results mutually confirm the validity
of the analytic formulae and of the DMRG results.

For $U\neq W$, the single-particle gap extrapolates to its value
in the thermodynamic limit with corrections of the order $1/L$.
At the Mott transition point, $U=W= 1$, the finite-size corrections
are of the order $1/\sqrt{L}$.

\subsection{Two-particle gap}

\begin{table}[b]
  \begin{center}
  \begin{tabular}[t]{|l|l|l|l|}\hline
    $L$ & $\Delta_2(0.5)$ & $\Delta_2(1)$
    & $\Delta_2(2)$  \\
    \hline
    4\vphantom{\Large $A^A$} & 0.866025403784
    & 1.50000000000   & 3.23205080757\\
    6 & 0.597095949159
    & 1.14982991426   & 2.86085856499\\
    8 & 0.457106781187
    & 0.957106781187  & 2.66421356237\\
    16 & 0.237372435696
    & 0.625000000000   &2.34974487139\\
    32 & 0.121516994375
    & 0.416053390593   &2.18053398875\\
    64 & 0.061580085890
    & 0.281250000000   &2.09191017178 \\
    128 & 0.031013203202
    &   0.192401695297 & 2.04640140640\\
    $\infty$ & 0 & 0 & 2\\\hline
  \end{tabular}
  \end{center}
  \caption{Two-particle gap $\Delta_2(U/W)\equiv \Delta_2(W,U,L)/W$
    of the $1/r$-Hubbard model
    with anti-periodic boundary conditions for $U/W=0.5,1,2$
    and $L=4,6,8,16,32,64,128$ at half band-filling.
    The DMRG reproduces the data with an accuracy of at
    least six digits.\label{tab:three}}
\end{table}

For the calculation of the two-particle gap, we again use
the energy formula~(\ref{eq:gsenergyfinite}). We thus find
\begin{equation}
\mu_2^-= \frac{U+W}{2}-\frac{W}{L}
  -\frac{1}{2} \sqrt{(W-U)^2+4WU/L} \; .  
\end{equation}
because only the energy difference of the last two sites remains in the
difference in the ground-state energies for $N=L$ and $N=L-2$ particles on $L$ sites.
Thus, from eq.~(\ref{eq:defDelta2}) we find
\begin{equation}
  \Delta_2(L;U) = U-W+\frac{2W}{L} +\sqrt{(W-U)^2+\frac{4WU}{L}} \;,
  \label{eq:Delta2analyt}
\end{equation}
which reduces to
\begin{equation}
  \Delta_2(U) = U-W+|W-U|=2 \Delta_1(U)
  \label{eq:Delta2neededlater}
\end{equation}
in the thermodynamic limit, as expected.
Some values for finite system sizes are collected in table~\ref{tab:three}.
The DMRG reproduces the values
with an accuracy of at least six digits.
Again, these results mutually confirm the validity
of the analytic formulae and of the DMRG results.

\begin{figure}[t]
 \includegraphics[width=8cm]{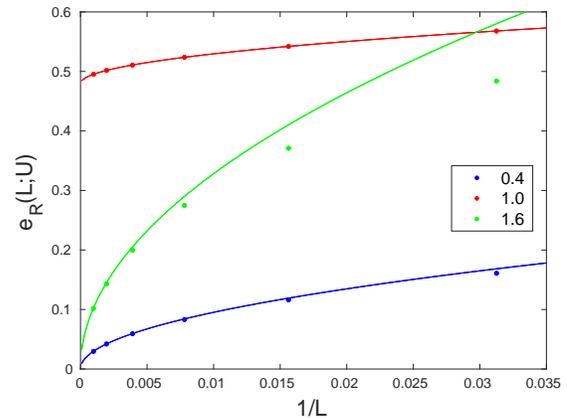}
 \caption{Effective repulsive energy $e_{\rm R}(L;U)$
   of two holes in the half-filled ground state,
   eq.~(\ref{eq:defdeltaeR}),
    multiplied by $\sqrt{L}$
    as a function of inverse system size ($L=32,64,128,256,512,1024$)
    for the $1/r$-Hubbard model
    for $U=0.4$, $U=U_{\rm c}=1$, and $U=1.6$. The lines result from
    eq.~(\ref{eq:eRnoncritical}) for $U=0.4$ and $U=1.6$,
    and from eq.~(\ref{eq:eRcritical}) for $U=1$.\label{fig:energyR}}
\end{figure}

In Fig.~\ref{fig:energyR} we show the effective
repulsive energy of the two holes confined to $L$ sites, eq.~(\ref{eq:defdeltaeR}).
Away from the transition, $e_{\rm R}(L;U)\sim 1/L$ which is characteristic for a
two-particle repulsion of finite range,
\begin{eqnarray}
  e_{\rm R}(L;U<U_{\rm c})&\approx & \frac{2U}{1-U^2}\frac{1}{L} \; , \nonumber \\
  e_{\rm R}(L;U>U_{\rm c})&\approx & \frac{2U^2}{U^2-1}\frac{1}{L} \; .
  \label{eq:eRnoncritical}
\end{eqnarray}
It is only at the critical interaction, $U_{\rm c}=1$,
that the correlation length diverges which results in
$e_{\rm R}(L;U_{\rm c}=1)\sim 1/\sqrt{L}$. For this reason, we 
actually plot $\sqrt{L} e_{\rm R}(L;U)$ in Fig.~\ref{fig:energyR}
which extrapolates to a finite value in the thermodynamic limit
when $U=U_{\rm c}=1$,
\begin{equation}
 \sqrt{L} e_{\rm R}(L;U=U_{\rm c})\approx
  \left(2+4(2\sqrt{2}-1)\zeta(-1/2)\right)
  +\frac{1}{2\sqrt{L}} \; .
  \label{eq:eRcritical}
\end{equation}
In the derivation
of eqs.~(\ref{eq:eRnoncritical}) and~(\ref{eq:eRcritical}),
we used {\sc Mathematica}~\cite{Mathematica12} to perform the sums
and the expansion in $1/L$.
Numerically, $\left(2+4(2\sqrt{2}-1)\zeta(-1/2)\right)\approx 0.479581$.

\section{Momentum distribution}
\label{sec:momentumdistribution}

The momentum distribution cannot be calculated analytically
in general but can only be evaluated
perturbatively for small coupling to order $(U/W)^2$ and
for strong coupling to order $W/U$. DMRG, however,
provides $n_k$ for systems with up to
128 sites for all interaction strengths.

\subsection{Momentum distribution at weak coupling}

\subsubsection{Wave function in weak coupling}

As shown by Girndt and one of us,\cite{Florianmodelvariational}
see also Dzierzawa et al.,\cite{Dionysredo}
the Gutzwiller wave function~\cite{Gutzwiller1963}
\begin{equation}
  |\Psi_{\rm G}(g)\rangle = g^{\hat{D}} | {\rm FS}\rangle
  \label{eq:defGWF}
\end{equation}
reproduces the ground-state energy
of the $1/r$-Hubbard model~(\ref{eq:fullHubbardmodel})
at half band-filling to order $U^2$. Here, $| {\rm FS}\rangle$
is the paramagnetic Fermi-sea ground state
at $U=0$ and $g$ is a variational parameter
with $1\geq g > 0$ for $0\leq U<\infty$.
By construction, the variational state is exact for $U=0$  where $g=1$.

At half band-filling we have~\cite{Metzner1987}
\begin{eqnarray}
  \bar{d}(g) &=&
\frac{1}{L}  \frac{\langle \Psi_{\rm G}(g) | \hat{D}| \Psi_{\rm G}(g) \rangle }{
    \langle \Psi_{\rm G}(g) | \Psi_{\rm G}(g) \rangle } \nonumber \\
  &=& \frac{g^2}{2(1-g^2)^2} \left[ -(1-g^2) -\ln(g^2)\right]
\end{eqnarray}
for the average double occupancy and
\begin{eqnarray}
\bar{T}(g) 
&=&\frac{1}{L}
  \frac{\langle \Psi_{\rm G}(g) | \hat{T}| \Psi_{\rm G}(g) \rangle }{
    \langle \Psi_{\rm G}(g) | \Psi_{\rm G}(g) \rangle } \nonumber \\
  &=& -\frac{1}{4} -\left( \frac{g-1}{g+1}\right)
  \left(\frac{1}{4} -\bar{d}(g)\right)
  \end{eqnarray}
for the average kinetic energy (bandwidth $W=2\pi t \equiv 1$).
For general~$U$, the minimum of the variational energy
\begin{equation}
E_{\rm var}(g) = \bar{T}(g) +U \bar{d}(g)
\end{equation}
must be obtained numerically. 

For $U\ll 1$ and thus $1-g\ll 1$ we find analytically
using {\sc Mathematica}~\cite{Mathematica12}
\begin{equation}
g(U)= 1-U -\frac{U^2}{2}-\frac{U^3}{5} +\alpha U^4 +{\cal O}(U^5)
\label{eq:gofU}
\end{equation}
with $\alpha$ of the order unity. Therefore, the variational upper bound
on the exact ground-state energy from the Gutzwiller wave function is given by
\begin{equation}
  e_{0,{\rm var}}^{\rm opt}(U\ll 1)\approx -\frac{1}{4} +\frac{U}{4}-\frac{U^2}{12}
  +\frac{U^4}{240} +{\cal O}(U^6)
  \label{eq:Gutzweakcoupling}
\end{equation}
for weak interactions.
It reproduces the second-order term exactly and overestimates
the fourth-order term because
\begin{equation}
  e_0(n=1;U\leq 1)= -\frac{1}{4} +\frac{U}{4}-\frac{U^2}{12}
  \label{eq:e0halffillingweakcouopling}
\end{equation}
from eq.~(\ref{eq:ezeroTDL}).
Since the prefactor of the fourth-order term in eq.~(\ref{eq:Gutzweakcoupling})
is small, the relative error of the Gutzwiller estimate is below one percent for
$U\lesssim 0.8$.

\subsubsection{Momentum distribution in the Gutzwiller wave function}
\label{subsubsec:Gutzwillermomdis}

Kollar and Vollhardt~\cite{KollarVollhardt} derived an analytic expression
for the momentum distribution for the Gutzwiller wave function
with a Fermi sea where the $k\sigma$-states occupy the region $|k|\leq \pi/2$,
\begin{eqnarray}
  n_{0\leq k \leq \pi/2,\sigma}(g) &=& \frac{g^2+4g+1}{2(1+g)^2} \nonumber \\
  &&  +\frac{g^2}{(1+g)^2}\frac{4{\cal K}[x(g,k)]}{\pi\sqrt{(2-G)^2-(\tilde{k}G)^2}}
  \nonumber \; , \\
  x(g,k)&=& 
  \frac{G\sqrt{1-\tilde{k}^2}}{\sqrt{(2-G)^2-(\tilde{k}G)^2}}
  \; , \nonumber \\
  G&=& 1-g^2 \; , \nonumber \\
  \tilde{k}&=& \frac{2k}{\pi} \leq 1 \;, \nonumber \\
        {\cal K}(x) &=& \int_0^{\pi/2}\rmd \varphi \frac{1}{\sqrt{1-x\sin^2(\varphi)}}
        \; ,
        \label{eq:momdisGutzwiller}
  \end{eqnarray}
and $n_{\pi/2<k<\pi,\sigma}(g)=1-n_{\pi-k,\sigma}(g)$.
Due to inversion symmetry, we have
$n_{ -k,\sigma}(g)=n_{ k,\sigma}(g)$.
Note that ${\cal K}(x)$ is the complete elliptic integral of the first kind.
The argument $x(g,k)$ in eq.~(\ref{eq:momdisGutzwiller}) obeys
$0\leq x(g,k)<1$ for $0<g\leq 1$ and $0\leq \tilde{k}\leq 1$.

The jump in the momentum distribution at $|k|=\pi/2$
is given by~\cite{Metzner1987}
\begin{equation}
  q_{\sigma}(g)=\frac{4g}{(1+g)^2} \;.
  \label{eq:qfactorGutzwiller}
\end{equation}

\subsubsection{Momentum distribution for the 1/r-Hubbard model}

For the discontinuity of $n_k(U)$ at the Fermi wave vector,
the Gutzwiller wave function predicts
\begin{equation}
  q(U\ll 1)\approx 2-\frac{U^2}{2} +\frac{U^4}{20} + {\cal O}(U^6)
  \label{eq:Gutzwillerjumpprediction}
\end{equation}
when we insert eq.~(\ref{eq:gofU}) into eq.~(\ref{eq:qfactorGutzwiller}).
For general~$k$, we expand $n_k(g)$ in eq.~(\ref{eq:momdisGutzwiller})
for small~$U$. For the momentum distribution up to order $U^4$ we find
the Gutzwiller wave-function prediction 
\begin{eqnarray}
  n_{-\pi< p \leq 0}(U) &=& 2
  + 2U^2\biggl[-\frac{3}{16}+\left(\frac{k}{2\pi}\right)^2
    \biggr] \nonumber \\
  && +2U^4 \biggl[ \frac{49}{1280} -\frac{39}{40} \left(\frac{k}{2\pi}\right)^2
    + 9 \left(\frac{k}{2\pi}\right)^4    \biggr]
  \; ,
  \nonumber \\
  k&=& p+\frac{\pi}{2}
  \label{eq:nksmallUGutzwillerprediction}
\end{eqnarray}
for the $1/r$-Hubbard model. By particle-hole symmetry,
$  n_p(U) = 2-n_{-p}(U)$.
The approximation~(\ref{eq:nksmallUGutzwillerprediction})
works well for $U\lesssim 0.4$, for momenta away from the band edges and away from
the discontinuity at the Fermi wave vector.

Note that for the $1/r$ Hubbard model
the Fermi sea is in the region $-\pi<p<0$, i.e., it is shifted by $\pi/2$
with respect to the expressions in Sect.~\ref{subsubsec:Gutzwillermomdis}.
Therefore, we must replace $k$ in eq.~(\ref{eq:momdisGutzwiller}) using
the relation $k=p+\pi/2$.

\subsection{Momentum distribution at strong coupling}

At strong coupling and half band-filling, the $1/r$-Hubbard model
reduces to the Heisenberg model with $1/r^2$ exchange
(Haldane-Shastry model),\cite{HaldaneShastryHaldane,HaldaneShastryShastry}
whose exact ground state is the Gutzwiller projected half-filled Fermi sea
with $g=0$ in eq.~(\ref{eq:defGWF}). Since the spin correlations
for the Haldane-Shastry model are known exactly,\cite{Gebhard1987}
the momentum distribution of the $1/r$-Hubbard model
can be calculated analytically to first order in $1/U$.

\subsubsection{Wave function in strong coupling}

At $t(r)\equiv 0$, the ground state
of the $1/r$-Hubbard model~(\ref{eq:fullHubbardmodel})
is $2^L$-fold degenerate at half band filling
because each site can be occupied by either spin species,
\begin{equation}
\hat{D}| \varphi_n\rangle = 0 \; , \quad n= 1,2,\ldots 2^L\; .
\end{equation}
The degeneracy is not lifted in first order perturbation theory because
a single hopping process leads to a state with one double occupancy,
\begin{eqnarray}
  \langle \varphi_m | \hat{T}| \varphi_n\rangle &=& 0 \; ,
  \quad n,m= 1,2,\ldots 2^L\; ,
  \nonumber \\[3pt]
  \hat{D} \hat{T}| \varphi_n\rangle &=& \hat{T} |\varphi_n\rangle
  \; , \quad n= 1,2,\ldots 2^L\; .
  \label{eq:PTstates}
  \end{eqnarray}
Thus, the problem to be solved in second-order degenerate perturbation theory
is the diagonalization of a $2^L\times 2^L$ matrix with 
the entries
\begin{equation}
  \tilde{H}_{n,m} = \sum_{|R\rangle}\langle \varphi_n|
  \frac{\hat{T}|R\rangle\langle R|\hat{T} }{E_0^{(0)}-E_R^{(0)}}
  | \varphi_m\rangle \; .
\end{equation}
Using eq.~(\ref{eq:PTstates}) gives $E_0^{(0)}-E_R^{(0)}= -U$
for all $|R\rangle$ so that
\begin{equation}
\tilde{H}= \hat{P}_{D= 0}\left(-\frac{1}{U} \hat{T}^2\right) \hat{P}_{D= 0}
\label{eq:defHTilde}
\end{equation}
defines the effective spin model in the
subspace of no double occupancy.\cite{Anderson1959}

Let $|\Phi_0\rangle$ be the ground state of $\tilde{H}$,
\begin{equation}
\tilde{H}| \Phi_0\rangle = e_0(U) |\Phi_0\rangle
\end{equation}
with $e_0(n=1;U)\equiv e_0(U)= {\cal O}(1/U)$.
Then, according to (non-degenerate)
perturbation theory, the ground state of the
Hubbard model~(\ref{eq:fullHubbardmodel})
to first order in $1/U$ is given by
\begin{equation}
  |\Psi_0^{(1)}  \rangle =
      \left( 1 -\frac{1}{U} \hat{T}\right)
      |\Phi_0\rangle \; .
      \label{eq:groundstatelargeU}
\end{equation}
This can also be seen explicitly by applying $\hat{H}$ to $|\Psi_0\rangle$
in the subspaces of zero and one double occupancy while
noticing that $e_0(U)$ is of the order $1/U$.

\subsubsection{Momentum distribution for the 1/r-Hubbard model}

Using the definition of the momentum
distribution~(\ref{eq:defmomentumdistribution}) and
the approximate ground state from eq.~(\ref{eq:groundstatelargeU}),
we find for $\Delta n_k=n_k(n= 1;U)-1$
\begin{eqnarray}
  \Delta n_k^{(1)}&=& -\frac{1}{UL}\sum_{l\neq m}
  e^{{\rm i }k(l-m)} \langle \Phi_0 |
    \hat{T}
    \bigl(
    \hat{c}_{l,\uparrow}^+ \hat{c}_{m,\uparrow}^{\vphantom{+}}
    +
    \hat{c}_{l,\downarrow}^+ \hat{c}_{m,\downarrow}^{\vphantom{+}}
    \bigr)     | \Phi_0\rangle \nonumber \\
    &&
    -\frac{1}{UL}\sum_{l\neq m}
  e^{{\rm i }k(l-m)} \langle \Phi_0 |
    \bigl(
    \hat{c}_{l,\uparrow}^+ \hat{c}_{m,\uparrow}^{\vphantom{+}}
    +
    \hat{c}_{l,\downarrow}^+ \hat{c}_{m,\downarrow}^{\vphantom{+}}
    \bigr)\hat{T}
    | \Phi_0\rangle \nonumber \\
    &= &
    -\frac{1}{UL}\sum_{l\neq m}t(m-l)
    e^{{\rm i }k(l-m)} \langle \Phi_0 | \nonumber  \\
    &&   \hphantom{-\frac{1}{UL}\sum_{l\neq m}}
    \left(1/2+ \hat{S}^{z}_m\right)\left(1/2- \hat{S}^{z}_l\right)
    -\hat{S}_l^-\hat{S}_m^+
    \nonumber \\
    && \hphantom{-\frac{1}{UL}\sum_{l\neq m}}
            -\hat{S}_m^-\hat{S}_l^+ +
   \left(1/2- \hat{S}^{z}_m\right)\left(1/2 + \hat{S}^{z}_l\right)\nonumber\\
 && \hphantom{-\frac{1}{UL}\sum_{l\neq m}}            
    | \Phi_0\rangle \; ,
    \label{eq:nklargealmostdone}
\end{eqnarray}
where we used that $\hat{n}_{m,\uparrow}= 1/2+ \hat{S}_m^z$ and
$\hat{n}_{m,\downarrow}= 1/2- \hat{S}_m^z$
in the subspace of zero double occupancy at half filling.
Equation~(\ref{eq:nklargealmostdone}) can be further simplified to
\begin{eqnarray}
  \Delta  n_k^{(1)}
  &=& -\frac{1}{U}\sum_{r= 1}^{L-1}
  t(r)e^{-\rmi k r} \nonumber \\
&&  + \frac{4}{U} \sum_{r= 1}^{L-1}   t(r)e^{-\rmi k r} \frac{1}{L} \sum_{l= 1}^L
  \langle \Phi_0 | \hat{\rm \bf S}_{r+ l}\cdot \hat{\rm \bf S}_l
  | \Phi_0\rangle \;.
\end{eqnarray}
We introduce the $z$-component of the spin-spin correlation function,
\begin{equation}
    C^{zz}(r) = 
      \frac{1}{L} \sum_{l= 1}^L\langle \Phi_0 |
      \hat{S}^z_{r+ l}\hat{S}^z_l
  | \Phi_0\rangle \; ,
\end{equation}
and use spin-rotation symmetry to arrive at
\begin{equation}
\Delta  n_k^{(1)}
  = -\frac{\epsilon(k)}{U}+ \frac{12}{U} \sum_{r= 1}^{L-1}
  t(r)e^{-\rmi k r} C^{zz}(r) 
  \label{eq:nofkfinallargeU}
\end{equation}
as our result to order $1/U$.

In the thermodynamic limit, the spin correlation function
is known for all distances,\cite{Gebhard1987}
\begin{equation}
C_{\rm HS}^{zz}(r)= \frac{(-1)^r}{4\pi r} \Si(\pi r) \; ,
\end{equation}
where
\begin{equation}
\Si(x) = \int_0^x \rmd  t\frac{\sin(t)}{t}
\end{equation}
is the sine integral.
In eq.~(\ref{eq:nofkfinallargeU}) this gives after a short calculation
\begin{equation}
  n_k(n=1;U\gg 1)
    =  1- \frac{k}{2\pi U}+ \frac{3k}{2\pi U} \ln \left|\frac{k}{\pi}\right|
  + {\cal O}\left(\frac{1}{U^2}\right)
  \label{eq:nofkLRHubbardmodelfinal}
\end{equation}
with the bandwidth $W= 2\pi t\equiv 1$ as energy unit.

In eq.~(\ref{eq:nofkLRHubbardmodelfinal}) we note the fact that
the derivative of the momentum distribution
is logarithmically divergent at $k= 0$. This is a consequence of the
long-range electron transfer.

\subsection{Momentum distribution for finite system sizes}

\begin{figure}[t]
  \begin{center}
\includegraphics[width=8cm]{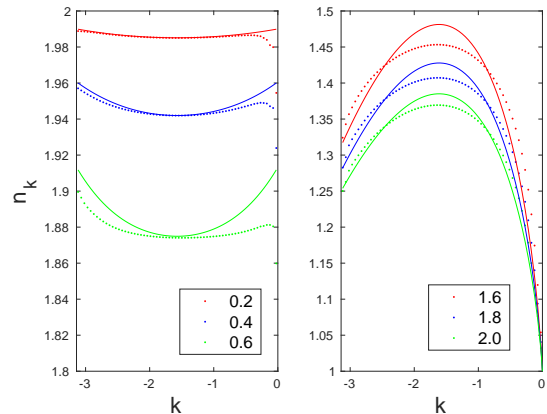}
  \end{center}
  \caption{Momentum distribution $n_k(L;U)$ for the $1/r$-Hub\-bard model
    for $U/W=0.2,0.4,0.6$ in the metallic phase (left), and
    for $U/W=1.6,1.8,2.0$ in the insulating phase (right) for $-\pi<k<0$.
    We superimpose the results for
    the four system sizes $L=16,32,64,128$. Continuous lines
    in the metallic phase are the predictions
    from the Gutzwiller wave function~(\ref{eq:nksmallUGutzwillerprediction}).
    Continuous lines in the insulating phase
are the predictions
    from the strong-coupling
    expansion~(\ref{eq:nofkLRHubbardmodelfinal}).\label{fig:nkOneoverrHM}}
\end{figure}

DMRG permits the calculation of the momentum distribution
for general on-site interactions and finite system sizes~$L$.
In Fig.~\ref{fig:nkOneoverrHM} we show $n_k(L;U)$, the momentum distribution
for the $1/r$-Hubbard model, as a function of
$k_m(L)=(2m+1)\pi/L$, see eq.~(\ref{eq:FTofoperators}),
for $U/W=0.2,0.4,0.6$ in the metallic phase and for $U/W=1.6,1.8,2.0$
in the insulating phase.
Since we study system sizes $L=2^R$ with $R=4,5,6,7$, the $k$-points
never coincide for different $L$.
Therefore, we combine all $k$-points in one figure noticing
that the $1/L$-corrections to $n_k(L;U)$ are fairly small
on the scale of the figures, apart from the region around the Fermi energy
and the band edges.

For weak coupling, the Gutzwiller
result~(\ref{eq:nksmallUGutzwillerprediction}) provides a reliable description
of the momentum distribution for $U\lesssim 0.4$, see the left part
of Fig.~\ref{fig:nkOneoverrHM}, 
apart from the region
close to the Fermi wave number $k_{\rm F}=0$ and away from the band edges
where perturbation theory
must break down because the model describes a Luttinger liquid and not
a Fermi liquid, as presumed in perturbation theory
around the Fermi-gas ground state.
Therefore, the perturbative result
for the jump discontinuity~(\ref{eq:Gutzwillerjumpprediction}) is not useful.

For strong coupling, the perturbative result~(\ref{eq:nofkLRHubbardmodelfinal})
applies (semi-)quantitatively for $U\gtrsim 1.6$
with small deviations around $|k|=\pi/2$,
see the right part of Fig.~\ref{fig:nkOneoverrHM}.
The comparison confirms the validity of the DMRG approach and permits
to set the limits for the applicability of the perturbative expressions.

\begin{figure}[t]
  \begin{center}
\includegraphics[width=8cm]{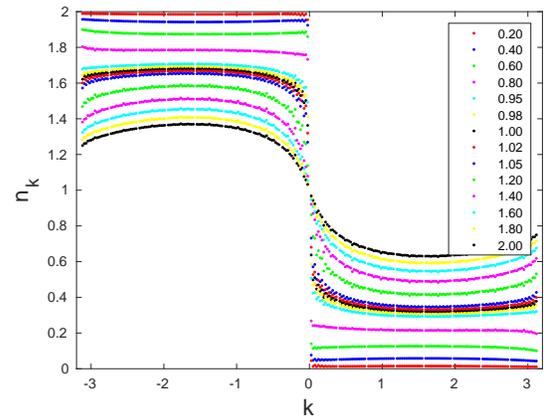}
  \end{center}
  \caption{Momentum distribution $n_k(L;U)$ for the $1/r$-Hub\-bard model
    for $U/W=0.6,0.8,1.0,1.2,1.4$.
    We superimpose the results for
    the four system sizes $L=16,32,64,128$.\label{fig:nkOneoverrHMinter}}
\end{figure}

In Fig.~\ref{fig:nkOneoverrHMinter} we show the momentum distribution also
for intermediate interaction strengths that cannot be accessed %
from perturbation theory. It is seen that
it poses a difficult problem to determine the size of the jump discontinuity
from data for finite system sizes.

\section{Mott-Hubbard transition from finite-size data}
\label{sec:locateMHtransition}

In generic one-dimensional Hubbard-type models, the Mott transition at half band-filling
occurs at $U_{\rm c}= 0^+ $ because the Umklapp scattering
is a relevant per\-tur\-ba\-tion.\cite{Giamarchi,Solyom3}
Concomitantly, it is exceedingly difficult for the Hubbard model with nearest-neighbor
electron transfer to identify the exponentially small gap
for small interactions.\cite{LiebWu,Gebhardbook}

In the $1/r$-Hubbard model, the gap is not exponentially small but opens
linearly at $U_{\rm c}=W$. It is interesting to see how
well the critical interaction can be determined from finite-size data
for the single-particle gap and for the momentum distribution.

\subsection{Finite-size data for the single-particle gap}
\label{subsec:finitesizedata}

The single-particle gap for all system sizes is given by
eq.~(\ref{eq:DelatfintieOoverrHM}). The analytical formula shows
that the gap scales as
\begin{eqnarray}
  \Delta_1(L;U\neq U_{\rm c})&=&\Delta_1(U)+a(U) \frac{1}{L}
  +{\cal O}\left(\frac{1}{L^2}\right)
  \; ,
  \\
    \Delta_1(L;U=U_{\rm c})&=&a(U_{\rm c})\left(\frac{1}{L}\right)^{1/2}+\frac{3}{4L}
    +{\cal O}\left(\frac{1}{L^{3/2}}\right)\nonumber 
    \label{finitesizegapOneoverrHM}
\end{eqnarray}
with
\begin{equation}
  a(U<U_{\rm c})=\frac{1}{1-U^2} \; , \;
  a(U>U_{\rm c})=1+\frac{U}{U^2-1} 
\end{equation}
and
\begin{equation}
    a(U_{\rm c})= 2(1-2\sqrt{2})\zeta(-1/2)\approx 0.76021\; .
\end{equation}

The analytic behavior of $\Delta_1(L;U)$ reflects the fact that the elementary
spin excitations of the
$1/r$-Hubbard model are gapless with a linear dispersion.
The elementary charge excitations also have a finite velocity
but with a finite gap in the insulating phase. At the critical interaction,
the charge velocity diverges proportional
to $1/\sqrt{L}$.\cite{GebhardGirndtRuckenstein}
In appendix~\ref{app:A} we perform the standard finite-size analysis
of the two-particle gap that does not lead to conclusive results
for $U_{\rm c}$. 

We follow a different approach and
combine the two cases in eq.~(\ref{finitesizegapOneoverrHM}) into
\begin{equation}
  \Delta_1(L;U)=\Delta_1(U)+ a(U)\left(\frac{1}{L}\right)^{\gamma(U)}
  \label{eq:gapextrapolationscheme}
\end{equation}
to find
\begin{equation}
  \gamma(U) = \begin{cases}
    \begin{array}{@{}lll@{}}
      1 & \text{for} & U\neq U_{\rm c} \\
      1/2 & \text{for} & U= U_{\rm c} =1\\
      \end{array}
  \end{cases} \; .
  \label{eq:defgammexponentexact}
\end{equation}
The prefactor $a(U)$ in eq.~(\ref{finitesizegapOneoverrHM})
diverges close to the transition,
\begin{equation}
  a(U\neq U_{\rm c}) \approx   \frac{1}{2} \frac{1}{|U-U_{\rm c}|} \; .
\end{equation}
Close to the transition, it thus requires system sizes $L\gg 1/|U-U_{\rm c}|$ to
reach the asymptotic regime where $\gamma(U)=1$ holds.

In numerical schemes such as the DMRG, we
perform calculations for systems with about one hundred sites
to keep the numerical effort limited.
To extract the gap from
finite-size data, we therefore
use the form~(\ref{eq:gapextrapolationscheme}) as our interpolation scheme.
We denote the numerically obtained values with the upper index ``$(L)$'',
e.g., $\Delta_1^{(L)}(U)$ for the extrapolated finite-size gap and
$\gamma^{(L)}(U)$ for the extrapolated exponent when using finite-size
data for chains with up to $L$ sites in the extrapolation.

\begin{figure}[t]
  \includegraphics[width=8cm]{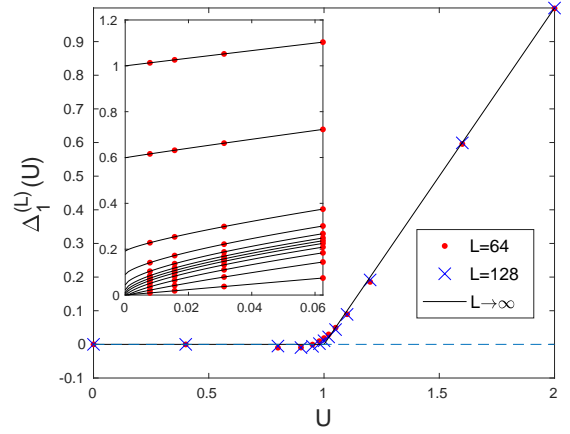}
  \caption{Single-particle gap $\Delta_1^{(L)}(U)$ for the $1/r$-Hubbard model
    as function of $U$, extrapolated from finite-size data with up to
    $L=64$ sites (points) and $L=128$ site (crosses), respectively.
    The continuous line is the exact result in the thermodynamic limit,
    eq.~(\ref{eq:gapTDL}). The inset shows the finite-size data
    and their extrapolation using eq.~(\ref{eq:gapextrapolationscheme})
    using the results for up to $L=128$ sites for
    $U/W=0,0.4,0.6,0.8,0.9,0.95,0.98,1,1.02,1.05,1.1,1.2,1.6,2$.
The intercept of the extrapolation curves with the ordinate
    defines the extrapolation estimate $\Delta_1^{(128)}$
    for the single-particle gap.\label{fig:singleparticlegap}}
\end{figure}

\begin{figure}[t]
  \begin{center}
      \includegraphics[width=8cm]{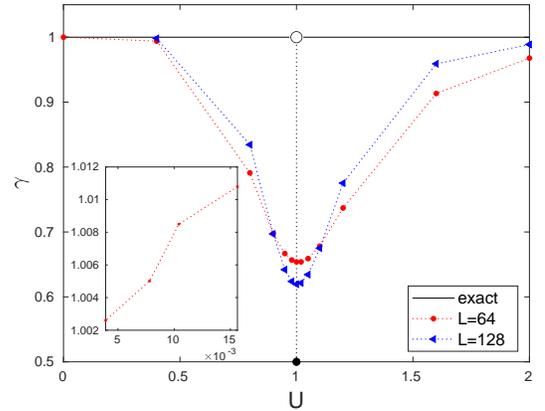}
  \end{center}
  \caption{Extrapolation exponents $\gamma^{(64)}(U)$ and $\gamma^{(128)}(U)$ 
    for the $1/r$-Hubbard model as a function of $U/W$.
The minimum of the curve determines $U_{\rm c}^{(L)}$.\label{fig:gammaofUinHM}}
\end{figure}

In Fig.~\ref{fig:singleparticlegap} we show the single-particle gap
$\Delta_1^{(L)}(U)$ for the $1/r$-Hubbard model
  as a function of~$U$ for $L=64,128$.
  In the inset,  we show the finite-size data for $L=16,32,64,128$ sites
  and the fit of the data
to the form~(\ref{eq:gapextrapolationscheme}).
It is seen that
the extrapolated data $\Delta^{(128)}(U)$ very well reproduce
the gap quantitatively but it is not clear how to
determine $U_{\rm c}$ accurately because the extrapolated 
curve $\Delta_1^{(L)}(U)$ is smooth and
cannot reproduce the kink in the analytical result
$\Delta_1(U)$ at $U=U_{\rm c}$.

For an accurate estimate of the critical interaction strength,
we must use a quantity that very sensitively depends on $U-U_{\rm c}$.
As can be seen from eq.~(\ref{eq:defgammexponentexact}),
the exponent $\gamma(U)$ is such a quantity
because it is one half
at the critical interaction
in comparison to $\gamma(U\neq U_{\rm c})=1$ for all other
interaction strengths, see eq.~(\ref{eq:defgammexponentexact}).
Of course, the isolated discontinuity at $U_{\rm c}$
cannot be reproduced from finite-size studies. However, %
$\gamma^{(L)}(U)$ retains its minimal
value at $U=U_{\rm c}^{(L)}$ that is close to $U_{\rm c}$, see 
Fig.~\ref{fig:gammaofUinHM}.

Apparently, the minimum of the curve $\gamma^{(L)}(U)$
can be determined very accurately.
A quadratic fit in the region $0.95\leq U \leq 1.05$ gives
$U_{\rm min}^{(64)}=1.011$ and $U_{\rm min}^{(128)}=1.005$.
At $L=128$, the deviation of $U_{\rm c}^{(128)}$ from the exact value
$U_{\rm c}=1$ is about five per mille.
When we linearly extrapolate the various values for $U_{\rm c}^{(L)}$
  for $L=64,96,128$, see the inset of Fig.~\ref{fig:gammaofUinHM},
  the exact result can be obtained with an accuracy of $2.5\cdot 10^{-4}$.
  
The gap exponent can be obtained with a similar precision.
As seen from eq.~(\ref{eq:gapTDL}), the gap opens linearly as a function of
the interaction, $\Delta_1(U>U_{\rm c})=(U-U_{\rm c})^{\nu}$ with $\nu=1$.
The fit of the gap data for $U\geq 1.02$ gives $\nu^{(128)}=1.003$
($\nu^{(64)}=0.987$),
within three (thirteen) per mille of the exact result.

\subsection{Finite-size analysis of the apparent discontinuity
  in the momentum distribution }

Next, we show that the apparent discontinuity of the momentum distribution
at the Fermi wave number cannot be used to determine
the critical interaction.

\begin{figure}[t]
  \begin{center}
      \includegraphics[width=8cm]{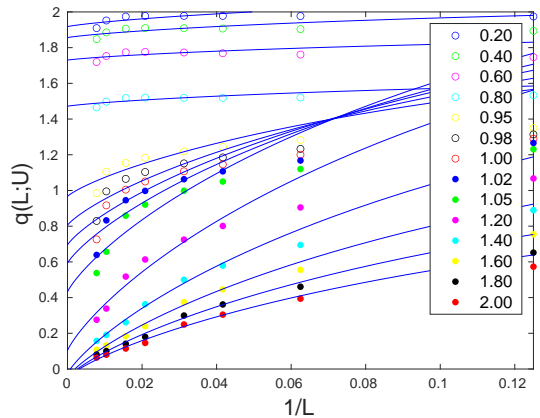}
  \end{center}
  \caption{Apparent jump discontinuity $q(L;U)$, eq.~(\ref{eq:apparentjump}),
    for the $1/r$-Hubbard model as a function of $1/L$
    for various $U/W$ and $L=8,16,24,32,48,64,96,128$.
    The lines use the fit function~(\ref{eq:fitjump}).
    \label{fig:jumpofnk}}
\end{figure}

In Fig.~\ref{fig:jumpofnk} we show the apparent discontinuity
of the momentum distribution,
\begin{eqnarray}
  q(L;U)&=&n_{-\pi/L}(L;U)-n_{\pi/L}(L;U) \nonumber \\
  &=& 2(n_{-\pi/L}(L;U)-1)
  \label{eq:apparentjump} \; ,
\end{eqnarray}
where we used particle-hole symmetry in the second step.
Inspired by the behavior for strong coupling, we use as our fit function
\begin{equation}
q(L;U) = q(U)+Q_1(U) \left(\frac{1}{L}\right)^{\beta(U)}\ln\left(\frac{1}{L}\right)
\label{eq:fitjump}
\end{equation}
for the extrapolation to extract $q(U)$.
The formula~(\ref{eq:fitjump}) can only apply when $\pi/L$ is rather close
to the Fermi edge so that we disregard $L=8,16$ in our fits.
The least-square optimization gives $|\beta-1|\ll 1$ for all~$U$.

As seen from Fig~\ref{fig:jumpofnk}, the extrapolation from $L=24,32,48,64,96,128$
sites
does not produce accurate results for the jump discontinuity.
For $U=U_{\rm c}$, the finite-size jump extrapolates to a sizable finite value
that persists down to $U=1.2$. For larger values of the interaction, $U\geq 1.2$,
the extrapolated gap becomes (slightly) negative.
Apparently, the jump discontinuity does not permit to determine
the critical interaction strength from system sizes up to $L=128$ sites.
System sizes of $L=10^3$ or even larger would be required to deduce $U_{\rm c}$
with a reasonable accuracy. Taking into account the scaling of the block entropy for
a fixed truncation error, 
these system sizes are beyond
our present computational capacities.

\begin{figure}[b]
  \begin{center}
          \includegraphics[width=8cm]{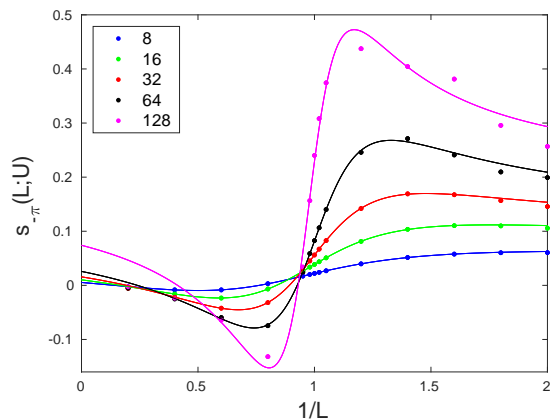}
      \end{center}
  \caption{Slope~$s_{-\pi}(L;U)$, eq.~(\ref{eq:slopedef}),
    as a function of~$U$ for $L=8,16,32,64,128$. The lines are fits to the
    Fano function~(\ref{eq:Fanofitfunction}).\label{fig:slopenB}}
\end{figure}

\subsection{Finite-size analysis of band-edge slope}

As seen from Figs.~\ref{fig:nkOneoverrHM} and~\ref{fig:nkOneoverrHMinter},
the momentum distribution has (local) extrema at the band edges
$k_{\rm B}=\pm \pi$. When we focus on the lower band edge, 
$n_{\rm \pi-\pi/L}$ displays a (local) minimum
in the insulating phase while there is a local maximum or minimum
in the metallic phase, depending on the system size.
Therefore, it is interesting to analyze the slope
of the momentum distribution at the band edge,
\begin{equation}
  s_{-\pi}(L;U)= \frac{L}{2\pi} \left[ n_{-\pi+\frac{3\pi}{L}}(L;U)
    -n_{-\pi+\frac{\pi}{L}}(L;U)  \right]  ,
   \label{eq:slopedef}
  \end{equation}
as a function of the system size and of the interaction~$U$.
In Fig.~\ref{fig:slopenB} we show the slope $s_{-\pi}(L;U)$
as a function of~$U$ for $L=8,16,24,32,48,64,96,128$.

The data resemble points on the curve of a Fano resonance.
In appendix~\ref{app:B} we provide some arguments under which
conditions a Fano resonance can show up in the slope $s_{-\pi}(L;U)$,
  \begin{eqnarray}
  s^{\rm Fano}_{-\pi}(L;U)&=&a_{-\pi}(L)  + \tilde{b}(L)
  \frac{[\Gamma(L) q_{\rm F}(L) +U-U_{\rm c}(L)]^2}{[\Gamma(L)]^2+[U-U_{\rm c}(L)]^2}
  \nonumber \\[3pt]
  \label{eq:Fanofitfunction}
  \end{eqnarray}
for $|U-U_{\rm c}|\ll U_{\rm c}$. 
For the five-parameter fit, we use the slope data
in the interval $0.4 \leq U \leq 1.6$,
from the metallic phase into the insulating phase.
In Fig.~\ref{fig:slopenB} we also display the slope $s^{\rm Fano}_{-\pi}(L;U)$
as a function of~$U$
for $L=8,16,32,64,128$.
The fits are very good, especially in the vicinity
of the critical interaction strength.

\begin{figure}[t]
  \begin{center}
      \includegraphics[width=8cm]{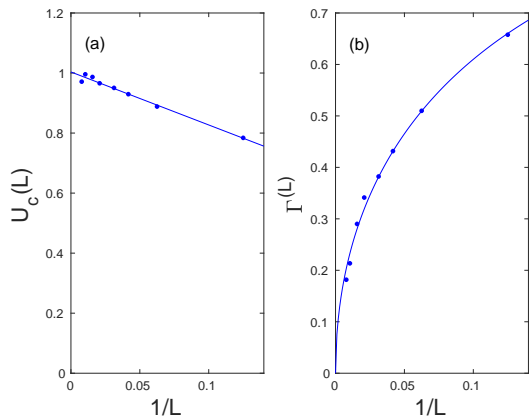}
  \end{center}
  \caption{(a)
    Critical interactions $U_{\rm c}(L)$, eq.~(\ref{eq:Fanofitfunction}),
    seen in the slope $s^{\rm Fano}_{-\pi}(L;U)$, eq.~(\ref{eq:slopedef}),
    as a function of inverse system size for $L=8,16,24,32,48,64,96,128$.
    The line is a linear fit in $1/L$.
(b) Width~$\Gamma^{(L)}$ of the resonance at $U=U_{\rm c}^{(L)}$,
    eq.~(\ref{eq:Fanofitfunction}),
    seen in the slope $s^{\rm Fano}_{-\pi}(L;U)$, eq.~(\ref{eq:slopedef}),
    as a function of inverse system size for $L=8,16,24,32,48,64,96,128$.
    The line is a quadratic fit in $1/\sqrt{L}$.
    \label{fig:Ucfromslope}\label{fig:Gammainnk}}
    \end{figure}

In Fig.~\ref{fig:Ucfromslope}(a)
we show the resulting values for $U_{\rm c}(L)$
as a function of $1/L$.
They linearly extrapolate to $U_{\rm c}(\infty)=1.004\pm 0.01$,
in agreement with the exact value $U_{\rm c}=1$ with an error of about one percent. To achieve a smaller error, we have to increase the system size and the
accuracy of the DMRG calculations for $L> 64$.
It is seen that, for the $1/r$-Hubbard model,
the critical interaction can be reliably determined from the slope of the momentum distribution at the lower band edge.

In Fig.~\ref{fig:Gammainnk}(b) we display the width of the
resonance~$\Gamma(L)$
in eq.~(\ref{eq:Fanofitfunction}). The width nicely extrapolates to zero
assuming a decay proportional to $1/\sqrt{L}$. As seen for the 
single-particle gap, eq.~(\ref{finitesizegapOneoverrHM}),
this scaling is characteristic for the critical interaction. In addition, the extrapolated value
confirms that there is a single-particle resonance at the band edge
in the thermodynamic limit at $U=U_{\rm c}$.

For completeness, we note that the Fano parameter is almost unity,
$q_{\rm F}(L\gtrsim 64) \approx 0.9\pm 0.1$. With the assumption $q_{\rm F}=1$ we
have in eq.~(\ref{eq:Fanofitfunction})
\begin{eqnarray}
  s^{{\rm Fano}, q=1}_{-\pi}(L;U)&=& a_{-\pi}(L)  + \tilde{b}(L)
   \label{eq:Fanofitfunctionq1} \\
  && +2 \tilde{b}(L)\Gamma(L)
  \frac{U-U_{\rm c}(L)}{[\Gamma(L)]^2+[U-U_{\rm c}(L)]^2} \; .
 \nonumber 
\end{eqnarray}
Since $\Gamma(L)\sim 1/\sqrt{L}$ for large system sizes and 
$\tilde{b}(L)\Gamma(L)$ must tend to a constant for large system sizes,
it is evident that $\tilde{b}(L\gg 1)\sim \sqrt{L}$, as we also confirm numerically. 
The values $a_{-\pi}(L)$ are negative and diverge for infinite system sizes,
$|a_{-\pi}(L)|\sim \sqrt{L}$, because $a_{-\pi}(L)+\tilde{b}(L)$ must remain
finite. 

Apparently, the slope $s_{-\pi}$ provides a useful method
to detect the transition in the $1/r$-Hubbard model.
It should be kept in mind that a singular behavior
of the slope of the momentum distribution $n_k$ at the band edge does not necessarily
prove the existence of a metal-insulator transition. We may argue, though, that
the occurrence of a single-particle bound state right at the band edge cannot
occur in the metallic or in the insulating phase but requires the peculiarities
of the transition point between both phases. 

\section{Conclusions}
\label{sec:conclusions}



In this work, we studied the one-dimensional Hubbard
model with a linear dispersion relation; the corresponding
electron transfer amplitudes decay proportional to the inverse
chord distance of two lattice sites on a ring (`$1/r$-Hubbard model').
Its exact spectrum was conjectured for all system sizes and
fillings.\cite{GebhardRuckenstein,Gebhardbook}
Using an efficient and accurate density-matrix renormalization group (DMRG) code,
we reproduced and thereby confirmed the conjectured energy formula
for $L\leq 128$ sites at half band filling (plus one or two particles),
with an accuracy of at least six digits for selected $U$-values.

The model provides an ideal case to study the Mott-Hubbard transition numerically 
because it lacks Um\-klapp scattering so that the critical interaction
occurs at a finite interaction strength,
$U_{\rm c}=W$, where $W$ is the bandwidth.
Moreover, the single-particle gap opens linearly above the transition,
$\Delta_1(U\geq W)=U-W$.
The critical properties of the spin and charge excitations for this model
are fairly simple,\cite{GebhardGirndtRuckenstein}
so that the finite-size scaling of the single-particle gap permits to locate the
critical interaction and the critical exponent with an accuracy of one per mille. 

DMRG also allows to calculate ground-state expectation values such as
the momentum distribution $n_k(L;U)$. For system sizes $L\leq 128$,
it is not possible to locate
the Mott transition from the apparent jump discontinuity at the Fermi wave vector.
Alternatively, we analyze the slope of the momentum distribution
at the band edge. It displays a critical behavior at the transition which reflects
the formation of a single-particle bound state at the band edge for $U=U_{\rm c}$.
Using the slope as a criterion for the Mott transition, the critical
interaction can be located only with an accuracy of one percent.
Note that  the occurrence of a single-particle bound state at the band edges
appears to be specific to the $1/r$-Hubbard model.


The main purpose of this work was to study
the Mott-Hubbard transition when it is not driven by Umklapp
scattering processes, and present alternative approaches to locate
quantum phase transitions in many-particle systems when conventional
extrapolations, e.g., for the gap, lead to inconclusive results,
see appendix~\ref{app:A}.
Moreover, in this work we demonstrated 
that the DMRG can be used efficiently to carry out the required numerical
simulations for large enough systems even for exotic models with long-range complex
electron-transfer amplitudes.

Our results open the way to study the Mott transition in one dimension
in the presence of long-range interactions.
It will be interesting to see how electronic screening in the metal, and its absence
in the insulator, modifies the Mott-Hubbard transition.
It is not yet clear whether or not the long-range Coulomb interactions alter
the Mott-Hubbard transition qualitatively, e.g.,
whether or not the gap opens continuously when the full screening problem
is addressed.\cite{Mottbuch}
We shall analyze this long-standing open question in a forthcoming publication.

\begin{acknowledgments}
  \"O.L.\ thanks the people at the Fachbereich Physik of the Philipps Universit\"at
  Marburg for their hospitality during the summer semester 2021.
  \"OL. has been supported by the Hungarian National Research,
  Development and Innovation Office (NKFIH) through Grants No.~K120569
  and No.~K13498, by the Hungarian Quantum Technology National Excellence
  Program (Project No.~2017-1.2.1-NKP-2017-00001)
  and by the Quantum Information National Laboratory of Hungary.
  \"O.L. also acknowledges financial support from the Alexander von Humboldt
  foundation and the Hans Fischer Senior Fellowship programme
  funded by the Technical University of Munich -- Institute for Advanced Study.
  The development of DMRG libraries has been supported by the Center for
  Scalable and Predictive methods for Excitation and Correlated phenomena (SPEC),
  which is funded as part of the Computational Chemical Sciences Program by the U.S.\
  Department of Energy (DOE), Office of Science, Office of Basic Energy Sciences,
  Division of Chemical Sciences, Geosciences, and Biosciences
  at Pacific Northwest National Laboratory. 
\end{acknowledgments}

\appendix

\section{Conventional gap extrapolation}
\label{app:A}

\begin{figure}[b]
  \begin{center}
     \includegraphics[width=8.6cm]{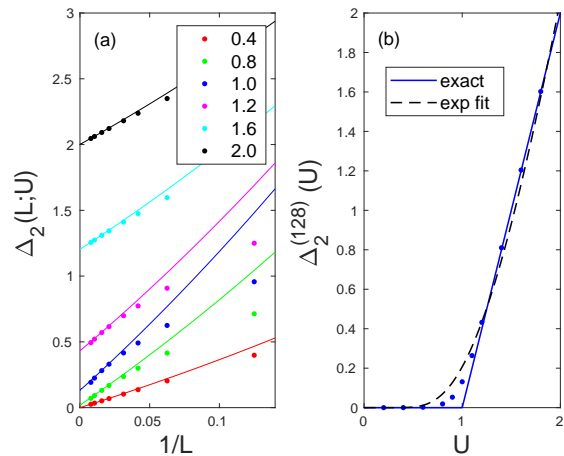}
  \end{center}
  \caption{(a) Two-particle gap $\Delta_2(L;U)$
    as a function of $1/L$ for 
    $L=8,16,32,48,64,96,128$ and $U=0.4,0.8, 1,1.2,1.6,2$.
    The lines are quadratic fits in the inverse system size,
    see eq.~(\ref{eq:myfitforDelta2}).
(b) Extrapolated two-particle gap $\Delta_2^{(128)}(U)$,
    for $U=0.2,0.4,0.6,0.8, 0.9,1,1.1,1.2,1.4,1.6,1.8,2$,
    in comparison with the analytic solution
    in the thermodynamic limit,
    see eq.~(\ref{eq:delta-tpgap-extrap}).
    The dashed black line shows the fit to eq.~(\ref{eq:KTsystem}).
    \label{fig:delta-tpgap-extrap}\label{fig:delta-tpgap}}
\end{figure}

For simplicity, we discuss the two-particle gap from eq.~(\ref{eq:Delta2analyt})
because it is given by a simple analytical formula,
\begin{equation}
  \Delta_2(L;U) = U-U_{\rm c}+\frac{2}{L} +\sqrt{(U_{\rm c}-U)^2+\frac{4U}{L}} \;.
  \label{eq:Delta2analytagain}
\end{equation}
As discussed in Sect.~\ref{subsec:gsprop}, it has the same analytical properties
as the single-particle gap. Eq.~(\ref{eq:Delta2analytagain}) shows that
the gap has a convergent Taylor expansion
in $1/L$ if $U\neq U_{\rm c}=W\equiv 1$.
Therefore, it seems natural to fit the gap for finite system sizes
to the function
\begin{equation}
  \Delta_2(L;U) \approx \Delta_2(U) + a(U)\frac{1}{L}+b(U)\frac{1}{L^2} \; .
  \label{eq:myfitforDelta2}
\end{equation}
In Fig.~\ref{fig:delta-tpgap}(a) we show the extrapolation
of the data for $L=8,16,24,32,48,64,96,128$ for $U=0.4,0.8,1,1.2,1.6,2$.

The extrapolation is seen to be very stable and $\Delta_2(U)$
can be determined reliably, apart from the
critical region where $\delta_U(L_{\rm max})
=|U-U_{\rm c}|\lesssim 2\sqrt{U_{\rm c}}/\sqrt{L_{\rm max}}$.
With $L_{\rm max}=128$ and $U_{\rm c}\approx 1$, we can expect deviations in the
region $0.8 \lesssim U \lesssim 1.2$.
Indeed, as seen from Fig.~\ref{fig:delta-tpgap-extrap}(b),
the extrapolation agrees very well
with the exact solution in the thermodynamic limit,
\begin{equation}
  \Delta_2(U\geq U_{\rm c})=\begin{cases}
    2(U-U_{\rm c}) & \text{for}\quad  U\geq U_{\rm c}=W\equiv 1\\
    0 & \text{for}\quad 0\leq U\leq U_{\rm c}
      \end{cases}\; ,
  \label{eq:delta-tpgap-extrap}
\end{equation}
see eq.~(\ref{eq:Delta2neededlater}),
outside the region $0.8 \lesssim U \lesssim 1.2$.

Inside this region, the continuous but sharp transition at $U_{\rm c}$
is smoothed out.
Therefore, it is rather difficult to derive the proper shape of the
two-particle gap in the thermodynamic limit.
For the present system sizes, even a fit to an exponential form
that applies to the exact gap of the standard one-dimensional Hubbard model
at small couplings,\cite{Essler}
\begin{equation}
  \Delta_2(U)^{\rm exp}=A \sqrt{U} \exp\left( -\frac{B}{U}    \right) \; ,
  \label{eq:KTsystem}
\end{equation}
appears to work in the region $0\leq U\leq 2$.
The fit with $A=10.28$ and $B=3.905$ is shown as a black dashed line
in Fig.~\ref{fig:delta-tpgap-extrap}(b).
This fit would incorrectly suggest $U_{\rm c}=0^+$ as in the standard Hubbard model.

In order to reduce the size of the critical region by a factor of ten, i.e.
down to $\delta_U=0.02$,
system sizes with $L_{\rm max}=10^4$ lattice sites would have to be
investigated, $\delta_U(10^4)=0.02$. Such system sizes cannot be treated numerically
with the required numerical accuracy
now or in the near future. For this reason, the conventional gap extrapolation
does not permit to determine $U_{\rm c}$ accurately from numerical data
for small system sizes.
Therefore, it is important to use the
extrapolation scheme introduced in Sect.~\ref{subsec:finitesizedata}
that permits an accurate estimate for $U_{\rm c}$ from data for up to
$L_{\rm max}=128$ sites.

\section{Fano resonance}
\label{app:B}

The Fano-Anderson model describes a localized state coupled to the
continuum.\cite{Fano1961,PhysRev.124.41} It provides a textbook example
for which the spectral function can be calculated analytically
using Green functions.\cite{Mahan}
For a Fano resonance at $\epsilon=\epsilon_0$, we have
\begin{equation}
  A_{\rm Fano}(b,\gamma,q_{\rm F},x)
  =b \frac{(\gamma q_{\rm F} +x)^2}{\gamma^2+x^2} \;,
  \label{eq:Fanoresdef}
\end{equation}
where $b$ is the strength of the resonance, $\gamma$ characterizes its width,
$q_{\rm F}$ is the Fano parameter, and $x=\epsilon-\epsilon_0$ denotes the deviation from
the resonance energy. For $q_{\rm F}=1$, the shape of the Fano resonance reduces to
\begin{eqnarray}
    A_{\rm Fano}(b,\gamma,q_{\rm F}=1,x) &=&
    b +2b \gamma \frac{x}{\gamma^2+x^2} \nonumber \\
    &=& b +2b\gamma {\rm Re}\left(\frac{1}{x+\rmi \gamma}\right)
        \; .\label{appeq:fitfuncslope}
\end{eqnarray}
This explains the counter-intuitive observation that a resonance
has the shape of the {\em real part\/} of a level
with a finite life-time~$\tau=1/\gamma$,
instead of its imaginary part.

To motivate the occurrence of a Fano resonance in the slope of the
momentum distribution, we assume that the spectral function contains
a part where frequency and momentum are related via a dispersion relation,
\begin{equation}
A(k,\omega)= A_{\rm reg}(k,\omega)+A_{\rm F}\left[\omega-v(k+\pi)/(2\pi)-f(U)\right]  \; .
  \end{equation}
Here, we focus on the lower band edge, $|(k+\pi)|\ll \pi$,
$v$ is the velocity of the excitations, and $f(U)$ is an unknown
function of the interaction that may also depend on the system size.
Now that at zero temperature~\cite{Mahan}
\begin{equation}
n_k= \int_{-\infty}^0 {\rm d}\omega A(k,\omega) \;, 
\end{equation}
we see that
\begin{eqnarray}
  \frac{\partial n_k}{\partial k}&=&
\frac{\partial   n_{k,{\rm reg}}}{\partial k}
-\frac{v}{2\pi } A_{\rm F}\left[-v(k+\pi)/(2\pi)-f(U)\right]  \; ,
\nonumber\\
\end{eqnarray}
where we used that $A(k,-\infty)=0$.\cite{Mahan}
Setting $k=-\pi$ we obtain
\begin{equation}
  s_{-\pi}=a_{-\pi}  -\frac{v}{2\pi } A_{\rm F}\left[-f(U)\right]  \; ,
  \label{eq:spianalytics}
\end{equation}
where we abbreviate $a_{-\pi}=(\partial   n_{k,{\rm reg}})/(\partial k)|_{k=-\pi}$.

Since a localized gapless state at the band edge
cannot exist but for $U=U_{\rm c}$, we may assume
\begin{equation}
  f(U)\approx f_0 (U-U_{\rm c})
  \label{appeq:AnsatzfofU}
\end{equation}
near the critical interaction.
We use the Ansatz~(\ref{appeq:AnsatzfofU})
and eq.~(\ref{eq:Fanoresdef}) in eq.~(\ref{eq:spianalytics})
and find after collecting all constants
\begin{equation}
  s_{-\pi}(L;U)=a_{-\pi}(L)  + \tilde{b}(L)
  \frac{[\Gamma(L) q_{\rm F}(L) +U-U_{\rm c}(L)]^2}{[\Gamma(L)]^2+[U-U_{\rm c}(L)]^2}
  \label{appeq:Fanofitfunction}
\end{equation}
for $|U-U_{\rm c}|\ll U_{\rm c}$, where we made explicit the dependency on the 
system size when the fit function~(\ref{appeq:Fanofitfunction})
is applied to finite-size data.



%

\end{document}